\documentclass[useAMS,usenatbib,usegraphicx]{mn2e}
\usepackage{amsmath}
\usepackage{amssymb}
\usepackage{multirow}

\def\simlt{\lower.5ex\hbox{$\; \buildrel < \over \sim \;$}}
\def\simgt{\lower.5ex\hbox{$\; \buildrel > \over \sim \;$}}

\newcommand{\bd}{\begin{displaymath}}
\newcommand{\ed}{\end{displaymath}}
\newcommand{\be}{\begin{equation}}
\newcommand{\ee}{\end{equation}}
\newcommand{\beqa}{\begin{eqnarray}}
\newcommand{\eeqa}{\end{eqnarray}}

\newcommand{\vbc}{v_{\rm bc}}
\newcommand{\LW}{{\rm LW}}
\newcommand{\cool}{{\rm cool}}
\newcommand{\tvir}{t_{\rm vir}}
\newcommand{\tmid}{t_{\rm mid}}

\title[21-cm signature during Lyman-Werner feedback]
{The 21-cm signature of the first stars during the Lyman-Werner
feedback era}
\author[Fialkov, Barkana, Visbal, Tseliakhovich \& Hirata]
{Anastasia Fialkov$^{1}$\thanks{E-mail: anastasia.fialkov@gmail.com},
Rennan Barkana$^{1}$, Eli Visbal$^{2, 3}$, Dmitriy Tseliakhovich$^{4}$,
\newauthor Christopher M. Hirata$^5$ \\ $^{1}$ Raymond and
Beverly Sackler School of Physics and Astronomy, Tel Aviv University,
Tel Aviv 69978, Israel \\ $^{2}$  Jefferson Laboratory of Physics, Harvard University, Cambridge, MA 02138, USA \\ $^{3}$  Institute for Theory $\&$ Computation, Harvard University, 60 Garden Street, Cambridge, MA 02138, USA \\ $^{4}$  California Institute of Technology,
M/C 249-17, Pasadena, California 91125, USA \\ $^{5}$ California
Institute of Technology, M/C 350-17, Pasadena, California 91125, USA }

\begin{document}
\pagerange{\pageref{firstpage}--\pageref{lastpage}} \pubyear{2012}
\maketitle

\label{firstpage}

\begin{abstract}

The formation of the first stars is an exciting frontier area in
astronomy. Early redshifts ($z \sim 20$) have become observationally
promising as a result of a recently recognized effect
\citep{Tseliakhovich:2010} of a supersonic relative velocity between
the dark matter and gas. This effect produces prominent structure on
100 comoving Mpc scales, which makes it much more feasible to detect
21-cm fluctuations from the epoch of first heating
\citep{Visbal:2012}. We use semi-numerical hybrid methods to
follow for the first time the joint evolution of the X-ray and
Lyman-Werner radiative backgrounds, including the effect of the
supersonic streaming velocity on the cosmic distribution of stars. We
incorporate self-consistently the negative feedback on star formation
induced by the Lyman-Werner radiation, which dissociates molecular
hydrogen and thus suppresses gas cooling. We find that the feedback
delays the X-ray heating transition by a $\Delta z \sim 2$, but leaves
a promisingly large fluctuation signal over a broad redshift
range. The large-scale power spectrum is predicted to reach a maximal
signal-to-noise ratio of S/N$\sim 3-4$ at $z\sim 18$ (for a projected
first-generation instrument), with S/N$>1$ out to $z \sim 22-23$.  We
hope to stimulate additional numerical simulations as well as
observational efforts focused on the epoch prior to cosmic
reionization.

\end{abstract}


\section{Introduction}

Observations of the redshifted 21-cm line of neutral hydrogen, planned
for the next decade, are expected to usher in a new era of direct
probing of the epoch of first stars. Though currently the main
observational focus is on the reionization epoch, there are
instruments hoping to observe the 21-cm signal from $z\sim
10-30$, e.g., LEDA \citep{Burns:2011},
DARE\footnote{http://lunar.colorado.edu/dare/}, and the SKA
\citep{Carilli:2004}.

The formation of the first stars is a relatively clean theoretical
problem, as they are formed in a metal-free environment via H$_2$
cooling \citep{Tegmark:1997, Machacek:2001, Abel:2002}. The radiation
produced by these first radiant objects changed the cosmic landscape
dramatically \citep{Madau:1997}. Three wavelength regimes of this
radiation are most important to consider: the Lyman-$\alpha$ photons
couple to the 21-cm line at high redshifts ($z\sim 30$) through the
Wouthuysen-Field effect \citep{Wouthuysen:1952, Field:1959}; X-ray
photons, produced by stellar remnants, heat the gas; and Lyman-Werner
(LW) photons dissociate molecular hydrogen, thus producing negative
feedback on star formation \citep{Haiman:1997} and decreasing the
heating rate.

The radiation spreads out to $\sim 100$ Mpc around each star, where
this finite effective horizon arises from redshift, time delay and
optical depth effects
\citep{Ahn:2009,21cmfast,Holzbauer:2011,Visbal:2012}. As star
formation progresses, the radiative backgrounds build up. Fluctuations
in the radiative backgrounds \citep{Barkana:2005,Pritchard:2006},
caused by the strongly fluctuating distribution of stars
\citep{Barkana:2004}, couple to the hyperfine transition of neutral
hydrogen and imprint fluctuations in the redshifted 21-cm
signal. Whereas Lyman-$\alpha$ coupling saturates at high redshifts
\citep{Holzbauer:2011, Visbal:2012},
the heating fluctuations couple to the 21-cm brightness temperature at
lower redshifts and thus are more interesting in terms of the
observational prospects. In particular, \citet{Visbal:2012} predict
detectible heating fluctuations from the first stars at $z = 20$ with
a distinctive signature of Baryon Acoustic Oscillations (BAOs)
imprinted by the supersonic relative (streaming) velocity between the
baryons and the dark matter \citep{Tseliakhovich:2010, Dalal:2010}.

In this paper we study the signature of the heating fluctuations
including the affect of relative velocity as in \citet{Visbal:2012},
and add for the first time a detailed three-dimensional calculation of
the inhomogeneous negative feedback by LW photons. We use the same
semi-numerical hybrid methods as in \citet{Visbal:2012} (based in part
on \citet{Tseliakhovich:2010} and \citet{21cmfast}) to build a
simulation where the stellar population and the radiative backgrounds
evolve simultaneously in time. The basic idea is to linearly evolve a
realistic sample of the Universe on large scales while using the
results of numerical simulations and analytical models to add in the
stars. For a detailed discussion of our computational methods, we
refer the interested reader to \citet{Visbal:2012}, in particular to
section S1 of this paper's Supplementary Information, and references
within. We use the standard set of cosmological parameters
\citep{WMAP7} along with an assumed star formation efficiency
(fraction of gas in star-forming halos that turns to stars) of $f_* =
10\%$, an X-ray photon efficiency of $10^{57}M_{\odot}^{-1}$ based on
observed starbursts at low redshift as in \citet{21cmfast}, and LW
parameters as explained in the next section.

\section{Incorporating the Negative Feedback}

The formation of the first stars via cooling of molecular hydrogen is
a highly non-linear process that can be mimicked by numerical
simulations, e.g., \citet{Abel:2002, Bromm:2002}. However, numerical
simulations in which primordial stars are created usually do not
consider the potentially fatal effect of the LW background on this
process. The negative feedback of the LW background on star formation
has been tested in the limited case of a fixed intensity $J_{\LW}$
\citep{Machacek:2001,Wise:2007,O'Shea:2008}. The feedback boosts the
minimal cooling mass, $M_{\rm cool}$, i.e., the mass of the lightest
halo in which stars can form, with the results of these simulations
well-described by the relation
\be \label{Mcool0} M_{\rm cool}\left(
J_{21}, z\right) = M_{\cool,0}(z)\times\left[1+6.96\left(4\pi
J_{21}\right)^{0.47}\right], \ee
where $J_{21} = J_{\LW}/(10^{-21}$erg
s$^{-1}$cm$^{-2}$Hz$^{-1}$sr$^{-1})$ in terms of the LW intensity
$J_{\LW}$; another common notation is the LW flux $F_{\LW} = 4\pi
J_{\LW}$. Here $M_{\cool,0}(z)$ is the value of the minimum cooling
mass in the standard case with no LW background.

This result is incomplete for two reasons. One is that it does not
account for the relative velocity $\vbc$, which has a strong impact on
the primordial star formation by (among other things) boosting the
minimum cooling mass \citep{Greif:2011,Stacy:2011,Fialkov:2011}. To
account for the velocity, we change $M_{\cool,0}(z)$ in
Eq.~(\ref{Mcool0}) to $M_{\cool,0}(z,\vbc)$, using the fit we
developed in \citet{Fialkov:2011} to the streaming-velocity
simulations (here we use their fit designed for Adaptive Mesh
Refinement (AMR) simulations as equation~(\ref{Mcool0}) was the result
of a fit to an AMR simulation). Thus, we combine two separate physical
phenomena, i.e., the relative motion and the LW flux, assuming that
they each have a fixed multiplicative effect on the minimum cooling
mass.  This simple ansatz for the dependence of $M_{\cool}$ on the two
parameters, $\vbc$ and $J_{21}$, should be checked by detailed
numerical simulation, which we hope to stimulate with this work. We
choose this multiplicative ansatz since both effects have an
independent effect on the minimal cooling mass. Therefore they likely
reinforce each other when they are both present. We note that while we
include the $\vbc$ and LW effects separately in Eq.~(1), our results
do account for the strong correlation between the velocity and the LW
flux (due to the effect of the velocity on star formation).

The second incompleteness of Eq.~(\ref{Mcool0}) is its validity only
in the case of a fixed background intensity during the formation of
the halo, whereas in reality the LW intensity is expected to rise
exponentially with time at high redshifts (e.g., see the Supplementary
Information in \citet{Visbal:2012}). Treating the intensity as fixed
at its final value would greatly overestimate the strength of the
feedback, since the cooling and collapse involved in star formation
should respond with a delay to a drop in the amount of H$_2$. For
instance, if the halo core has already cooled and is collapsing to a
star, changing the LW flux may not stop or reverse the collapse at
all, and certainly not immediately. Another indication for the gradual
process involved is that the simulation results can be approximately
matched \citep{Machacek:2001} by comparing the cooling time in halo
cores to the Hubble time (which is a relatively long timescale).
Though the relation in Eq.~(\ref{Mcool0}) is the best currently
available, more elaborate numerical simulations, which we again hope
to stimulate, are needed in order to find a more realistic dependence.
We overcome this limitation by using the above relation not with the
final value of $J_{21}$ at formation, but with the value at a mean,
characteristic time within the halo formation process.

The idea of looking at the flux at times well before virialization is
based on an analogy with the filtering mass defined in the
well-studied case of pressure \citep{Gnedin:1998,Gnedin:2000}. In the
latter case, the actual gas fraction in non-linear, virialized halos,
is close to the filtering mass, not the Jeans mass, and the filtering
mass is affected by the value of the Jeans mass at much earlier times.
The reason the gas fraction is affected by the Jeans mass at early
times is that the relationship between the gas temperature and the gas
density is very indirect; the temperature affects the pressure
gradient, and thus the acceleration, which affects the actual position
after a delay. In particular, if the temperature drops suddenly, the
gas that was far away from the halo center does not instantly fall
inside. In the case of cooling, there are the additional steps from
dissociation of H$_2$, which changes its abundance, through the
process of cooling which then affects the temperature. The cooling
history thus affects the distribution of gas. The LW flux rises
exponentially fast with time and was very small at early times. Thus,
the cooling was fast initially, and the gas cooled and started to
collapse. A sudden late rise in the flux may not be able to stop this
collapse.

Using the characteristic value of $J_{21}$ with a realistically large
uncertainty should suffice for our main goal of spanning the possible
range of the effect of $J_{\LW}$ and $\vbc$ on the 21-cm background
during the X-ray heating era.
Specifically, we consider two possible feedback strengths which we
refer to as ``weak'' and ``strong'' feedback. Namely, for halos
forming (i.e., virializing) at some time $\tvir$, we adopt the
effective LW flux $J_{21}$ in Eq.~(\ref{Mcool0}) as the LW flux in the
same pixel at an earlier time $\tmid$, i.e., at the midpoint of the
halo formation process. In order to obtain a realistically large range
of uncertainty, with the spherical collapse model in mind we either
assume that ``formation'' spans the beginning of expansion up to
virialization (i.e., $t=0$ to $t=\tvir$, giving
$\tmid=\frac{1}{2}\tvir$: weak feedback), or just the collapse stage
starting at turnaround (i.e., $t=\frac{1}{2}\tvir$ to $t=\tvir$,
giving $\tmid=\frac{3}{4}\tvir$: strong feedback). We also compare to
the limiting cases (shown in \citet{Visbal:2012}) of no feedback or
saturated feedback. The latter correponds to assuming that star
formation is only possible via atomic cooling; in this scenario, the
LW feedback is so efficient that H$_2$ is completely dissociated early
on and stars form in atomic cooling halos ($M_{\cool} \ga 3
\times 10^7 M_{\odot}$), as opposed to $M_{\cool}\la
10^6 M_{\odot}$. For reference, we also consider the no feedback case
without the streaming velocity, in order to assess the importance of
the velocity effect. For given parameters, at each redshift the cosmic
mean gas fraction in stars decreases in the different cases in the
order: no feedback no velocity, no feedback, weak feedback, strong
feedback, and saturated feedback (where all cases except the first
include the streaming velocity effect).

\begin{figure*}
\includegraphics[width=3.4in]{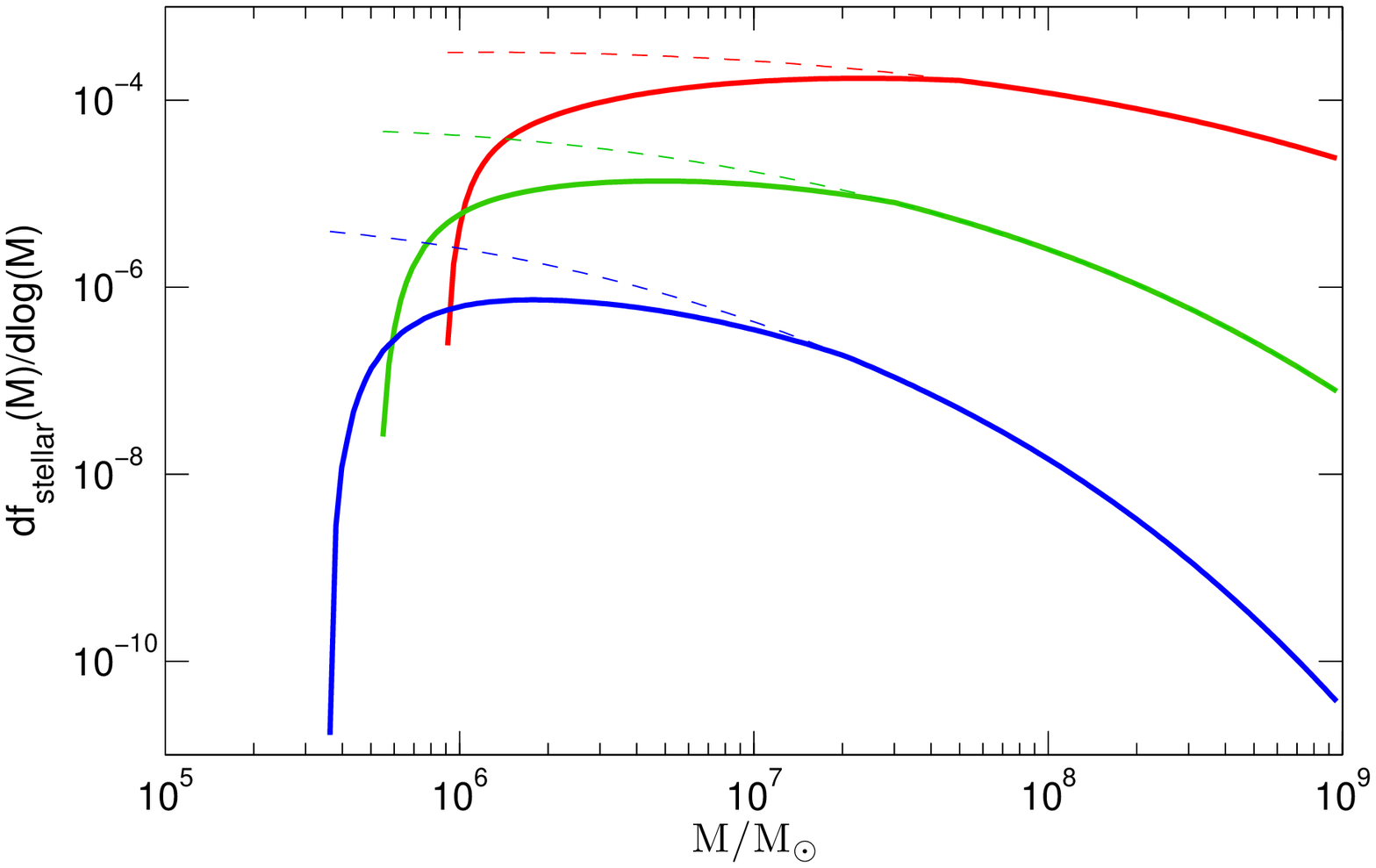}
\includegraphics[width=3.4in]{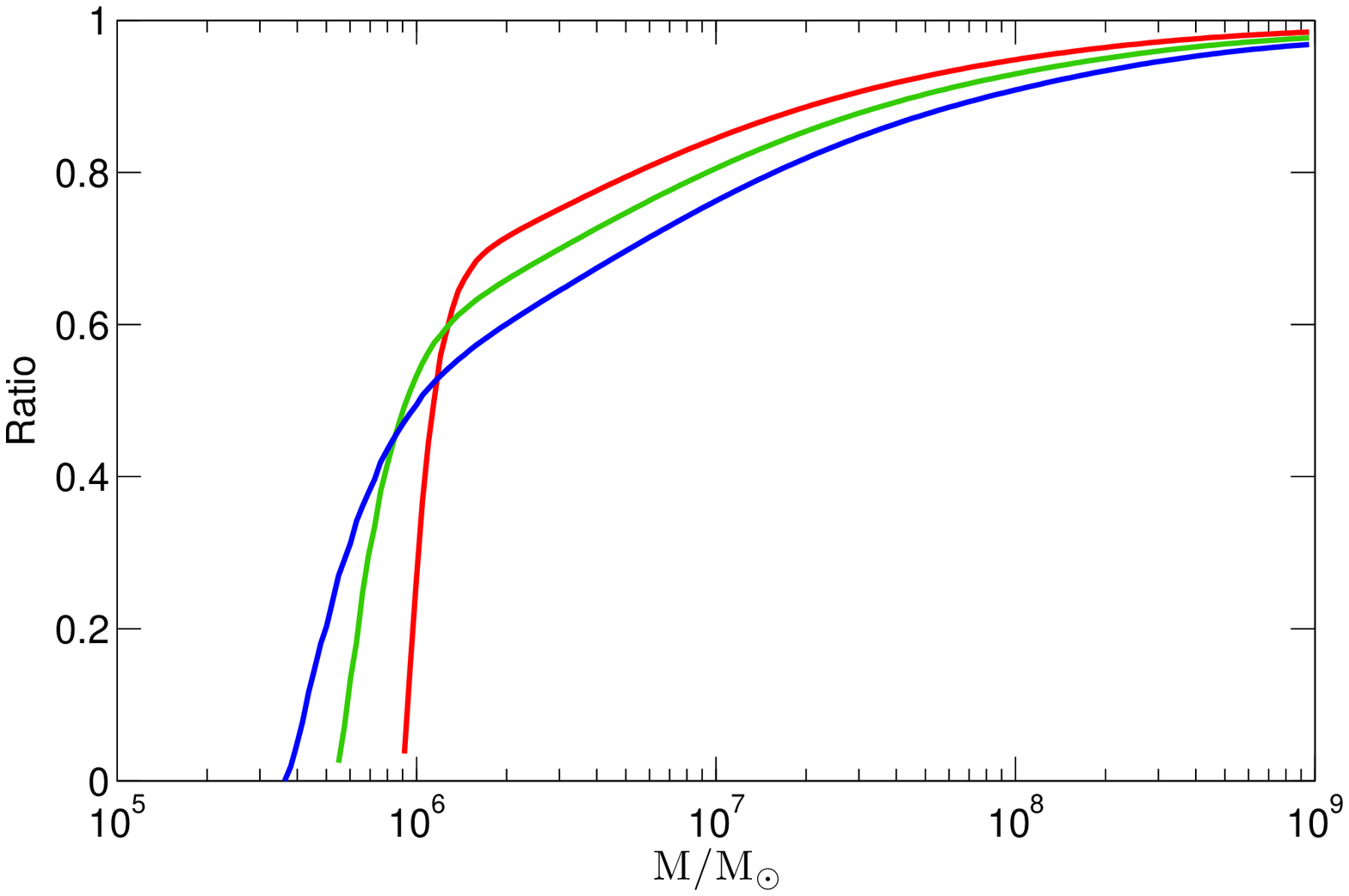}
\caption{Star-formation contribution and effect of velocities
versus halo mass (LW feedback not included). {\bf Left}: The
logarithmic contribution of each halo mass to the total fraction of
gas in stars (i.e., $d f_{\rm stellar}/d \log(M)$ averaged over the
distribution of $\vbc$), including the $\log(M)$ modulation in
Equation~(\ref{eq:f*}) (solid) or with the standard assumption of a
fixed efficiency with mass (dashed). We consider $z = 13.6$ (red),
$z = 19.6$ (green), and $z = 25.6$ (blue). {\bf Right}: The ratio of the
cosmic mean stellar fraction with $\vbc$ to the value without the
velocity effect, i.e., $<f_{\rm stellar}(M,\vbc)>_{\vbc} / f_{\rm
stellar}(M,0)$. We include Equation~(\ref{eq:f*}), and consider the
same redshifts as in the left panel.}
\label{fig:dfdm}
\end{figure*}

In our hybrid simulations we also incorporate two elements of the
astrophysics that make our calculations more complete and
accurate. One aspect is that we include a gradual low-mass cutoff for
star formation, rather than a sharp cutoff at $M_{\rm cool}$ as we
(and others) have previously assumed. Since the cooling rate declines
smoothly with virial temperature, a smooth cutoff is expected
physically, and indeed \citet{Machacek:2001} found that the fraction
of highly-cooled, dense gas in their simulated halos is well described
as being proportional to $\log(M/M_{\rm cool})$. Since this is the gas
that can participate in star formation, we incorporate this by
generalizing the star-formation efficiency to include a dependence on
halo mass, $f_*(M)$. We assume our standard efficiency of $f_* = 10\%$
for $M \ge M_{\rm atomic}$, where $M_{\rm atomic}$ is the minimum mass
for atomic cooling ($\sim 3 \times 10^7 M_\odot$ but $z$-dependent).
In order for $f_*(M)$ to be a continuous function, we thus set
\be \label{eq:f*}
f_*(M) = \left\{ \begin{array}{ll}
f_* & \mbox{if $M \ge M_{\rm atomic}$} \\
f_* \frac{\log(M/M_{\rm cool})} {\log(M_{\rm atomic}/M_{\rm cool})}
& \mbox{if $M_{\rm cool} < M < M_{\rm atomic}$} \\
0 & \mbox{otherwise.}
\end{array} \right.
\ee

As shown in Figure~\ref{fig:dfdm} (left panel), the standard
assumption of constant $f_*$ makes halos with masses near $M_{\rm
cool}$ dominate the cosmic star formation rate, particularly at the
highest redshifts. Our more realistic model significantly reduces the
overall star formation rate (by a factor of 2.0 in the example shown
at $z=19.6$) and shifts the peak of the contribution to star formation
to a higher mass ($8.7\times M_{\rm cool}$ at $z=19.6$). Also shown in the
Figure (right panel) is the overall effect of the relative velocity
$\vbc$ broken down by mass. Since the velocity effect on halos is made
up of three distinct effects, with two of them dominant
\citep{Fialkov:2011}, the dependence on halo mass shows two separate
regimes. Near the cooling mass (and up to a factor of $\sim 2$ above
it), the velocity effect is very strong and also strongly dependent on
$M$, mainly due to the boosting of the cooling mass in regions with a
high $\vbc$. At higher masses, however, the velocity effect is weaker
and only changes rather slowly with halo mass, mainly due to the
suppression of the halo abundance. A small but non-negligible effect
remains even well above $M_{\rm atomic}$. Since the velocity effect is
strongest at the low-mass end (right panel), the shifting of the star
formation towards higher masses (left panel) reduces somewhat the
overall influence of the supersonic streaming velocities. Since the LW
feedback also affects low masses first, the $\log(M)$ modulation
delays the LW feedback.

The second aspect of realistic astrophysics that we incorporate is
more directly related to the LW feedback. The LW photons emitted by
each source are absorbed by hydrogen atoms as soon as they redshift
into one of the Lyman lines of the hydrogen atom; along the way,
whenever they hit a LW line they may cause a dissociation of molecular
hydrogen. Some previous papers \citep{Ahn:2009,Holzbauer:2011} assumed
a flat stellar spectrum in the LW region and a flat absorption profile
over the LW frequency range. We incorporate the expected stellar
spectrum of Population III stars from \citet{Barkana:2005} (based on
\citet{Bromm:2001}), which varies in the LW region typically by a few
percent but up to $17\%$. More importantly, we explicitly include the
full list of 76 relevant LW lines from \citet{Haiman:1997}. We
summarize the results with $f_{\rm LW}$, the relative effectiveness of
causing H$_2$ dissociation via stellar radiation. Specifically it is
the ratio between the dissociation rate of molecular hydrogen and the
naive total stellar flux (i.e., calculated without any absorption and
integrated over all wavelengths), normalized to unity in the limit of
zero source-absorber distance. This quantity is simply a function of
source-absorber distance at each redshift under the simplifying
assumption of a universe at the mean density. This assumption follows
our approach for X-rays (as in 21CMFAST, \citet{21cmfast}), and should
be sufficiently accurate since the strong bias of star-forming halos
at these high redshifts implies that fluctuations in star formation
(which drive the 21-cm fluctuations) are much larger than the
fluctuations in the underlying density. Thus, give our assumed stellar
spectrum we can pre-calculate $f_{\rm LW}$ and include this as an
effective optical depth that is spherically symmetric around each
source. Any such symmetric effect is easily incorporated within the
numerical method of 21CMFAST which uses Fourier transforms to rapidly
perform averages over spherical shells.  

\begin{figure}
\includegraphics[width=3.4in]{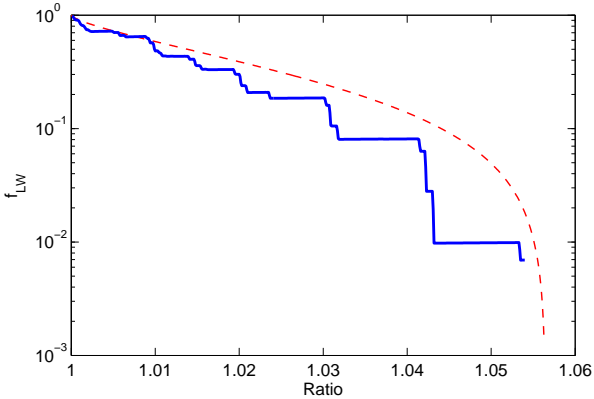}
\caption{The relative effectiveness of causing H$_2$ dissociation in
  an absorber at $z_a$ due to stellar radiation from a source at
  $z_s$, shown versus the ratio $R\equiv (1+z_s)/(1+z_a)$ (solid). For
  comparison we show $f_{mod}$, a commonly used approximation from
  \citet{Ahn:2009} (dashed). Both functions are normalized to unity at
  $R=1$. (There was also a $1.45\%$ normalization difference after we
  carefully normalized as in \citet{Machacek:2001}, since we use their
  results for the LW feedback.)}
\label{fig:fmod}
\end{figure}

Figure~\ref{fig:fmod} shows $f_{\rm LW}$ versus the absorber-source
distance; we parametrize this distance in terms of the absorber-source
scale-factor ratio $R$, since $f_{\rm LW}$ versus $R$ is independent
of redshift. Beyond the max shown $R=1.054$ (which corresponds to 104
comoving Mpc at $z=20$), {\bf $f_{\rm LW}$} immediately drops by five
orders of magnitude. The Figure shows that LW absorption is poorly
approximated as being uniform in frequency. In reality, emission from
distant sources is absorbed more weakly. For example, in one of our
main examples in the following section (i.e., our strong LW feedback
case including velocities), assuming a uniform spectrum and absorption
profile at $z=20$ would imply that typically, an atom receives $50\%$
of its LW flux from sources out to a distance of 18.9 Mpc, $80\%$ from
up to 42.2 Mpc, and $90\%$ from up to 55.8 Mpc. Our more accurate {\bf
  $f_{\rm LW}$} reduces these numbers to 14.4, 33.0, and 46.2 Mpc,
respectively.  The accurate {\bf $f_{\rm LW}$} reduces the overall LW
intensity by $\sim 20\%$ (thus delaying the LW feedback), and makes it
more short-range (i.e., local) and variable.

\section{Results}

\subsection{Mean Evolution}

The relative velocity amplifies heating fluctuations in the 21-cm
power spectrum, making it possible to observe the BAO in the first
stars \citep{Visbal:2012,McQuinn:2012}. In this section we consider
the effect of the negative LW feedback on this exciting observational
prospect. Our simulation evolves a realistic sample of the universe
from $z = 60$, roughly when the first stars turn on \citep{Naoz:2006,
Fialkov:2011}. We follow a box that is 384 Mpc on a side (all
distances comoving), with a pixel size of 3 Mpc. The X-ray and LW
backgrounds in each pixel are made up of contributions by stars
located within the corresponding effective horizons. Since we focus on
the era after Lyman-$\alpha$ coupling but prior to significant
reionization, the 21-cm brightness temperature (relative to the cosmic
microwave background (CMB) temperature $T_{\rm CMB}$)\cite{Madau:1997}
\begin{equation}
T_{\rm b} = 40 (1+\delta) \left( 1-\frac{T_{\rm CMB}} {T_{\rm K}}
\right) \sqrt{\frac{1+z}{21}} ~{\rm mK} \label{eq:Tb}
\ ,\end{equation}
where $\delta$ is the gas overdensity and $T_{\rm K}$ its kinetic
temperature.

\begin{figure}
\includegraphics[width=3.4in]{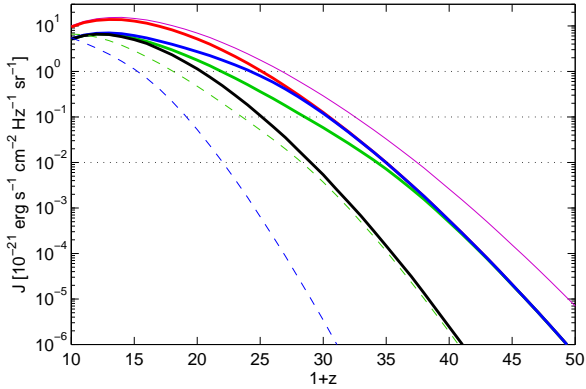}
\caption{The actual LW intensity
(solid lines) and the effective LW intensity for feedback on star
formation (dashed; shown only for the two realistic feedback
cases). We show the cosmic mean intensity (i.e., averaged over our
box) versus $1+z$ in the following cases: no feedback no $\vbc$
(purple), and with $\vbc$: no feedback (red), weak feedback (blue),
strong feedback (green), and saturated feedback (black).}
\label{fig:JLW}
\end{figure}

The negative feedback suppresses star formation on average, which
leads to a slower rise of the radiative backgrounds, and delays
various milestones of the star formation history. Figure~\ref{fig:JLW}
shows the rise of the mean LW flux with time. Comparing the two
no-feedback cases, we see that the streaming velocity has a large
overall suppression effect at high redshifts, which reaches about an
order of magnitude at $z>40$ but becomes quite small at $z < 15$.
Feedback can potentially be very strong at high redshifts (as
indicated by the saturated feedback case), but in practice the LW
feedback is expected to begin only when the effective flux reaches a
level of $J_{21} \sim 10^{-5}$; this happens at around redshift 30
(weak feedback) or 40 (strong). In both realistic feedback cases, the
LW feedback effectively saturates at $z \sim 10$.

We can easily understand why the two feedback cases converge with
time. Initially, the effective LW flux for star formation [i.e.,
$J_{21}$ in Eq.~(\ref{Mcool0})] is much higher at a given $z$ for the
strong feedback case (which assumes a less delayed value, closer to
the value of $J_{21}$ at $z$). The strong resulting feedback leads to
a slower rise of the actual $J_{21}$ and thus, eventually, also of the
effective $J_{21}$, compared to the weak feedback case. Therefore, the
effective $J_{21}$ in the weak feedback case gradually catches up with
the strong feedback case. Also important is that the rate of increase
of the flux naturally slows with time (i.e., the curves flatten),
since star-forming halos become less rare (i.e., they correspond to
less extreme fluctuations in the Gaussian tail of the initial
perturbations). The weak feedback case effectively looks back to
$J_{21}$ at an earlier time, when the rise was faster.

Figure~\ref{fig:JLW} tracks the rise of the LW flux through several
milestone values. A reasonable definition of the central redshift
$z_{\LW}$ of the LW transition is a mean effective intensity of
$J_{21}=0.1$, at which the minimum halo mass for cooling (in the
absence of streaming velocities) is raised to $\sim 2 \times 10^6
M_{\odot}$ due to the LW feedback. This is a useful fiducial mass
scale, roughly intermediate (logarithmically) between the cooling
masses obtained with no LW flux and with saturated LW flux. The
central range of the LW feedback transition can be defined by the
effective LW flux coming within an order of magnitude of its central
value, so that the minimum $M_{\cool}$ goes from $8 \times 10^5
M_{\odot}$ to $5 \times 10^6 M_{\odot}$ during this period.

\begin{figure}
\includegraphics[width=3.4in]{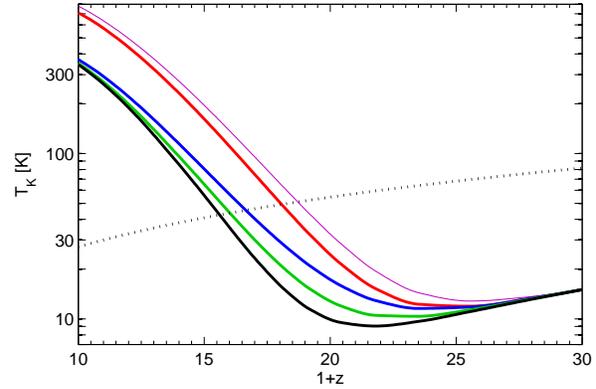}
\caption{The cosmic heating transition. We show the cosmic mean
gas kinetic temperature versus $1+z$ in the following cases: no
feedback no $\vbc$ (purple), and with $\vbc$: no feedback (red), weak
feedback (blue), strong feedback (green), and saturated feedback
(black). Also shown for comparison is the CMB temperature (dotted),
which crosses the gas temperature at $z_{\rm h}$.}
\label{fig:TK}
\end{figure}

Feedback also slows down the heating of the Universe
(Figure~\ref{fig:TK}). For example, the average heating rate at
redshift 20 for the weak, strong and saturated feedbacks are $55.9\%$,
$33.7\%$ and $19.1\%$ of the heating rate with no feedback (all
including the streaming velocity). As a result, the heating transition
is delayed. There are two possible natural definitions for this
transition, the standard more physical definition as the redshift
$z_{\rm h}$ when the mean gas temperature equals that of the CMB, and
the more observational (or 21-cm-centric) definition as the redshift
(which we denote $z_0$) at which the cosmic mean 21-cm brightness
temperature vanishes $\langle T_{\rm b} \rangle =0$. We consider both
definitions, but due to our focus on observational predictions, we
mostly use $z_0$.

In our simulation, $z_{\rm h} = 17.1$ and $z_0=16.6$ for the
no-feedback case (no feedback and no velocity gives $z_{\rm h} = 17.7$
and $z_0= 17.4$), while saturated feedback would delay these
milestones to $z_{\rm h}=14.6$ and $z_0 = 14.2$. The realistic
feedback cases are intermediate: $z_{\rm h}=15.7$ and $z_0=15.2$ for
the weak feedback case (with a LW transition centered at $z_\LW=19.2$,
and a central range of $z=22.0-15.2$), while $z_{\rm h}=15.0$ and
$z_0=14.6$ for strong feedback (with $z_\LW=23.6$, and a central range
of $z=28.3-18.1$). In every case, the LW transition starts very early,
and passes through its central redshift before the heating transition
(with a much bigger delay between the two transitions in the strong
feedback case). We note that if all the fluctuations were linear, then
we would find $z_{\rm h}=z_0$ identically. The difference of $\Delta z
= 0.4-0.5$ between them is an example of the effect of non-linear
fluctuations (plus, in this case, of the non-linear dependence of the
brightness temperature on the gas temperature). This shows that
analyses of this era based on linear theory can only give rough
estimates, and a hybrid simulation like ours is necessary in order to
properly incorporate the non-linear fluctuations in stellar density
and other derived quantities.

\subsection{Spatial Fluctuations}

\begin{figure*}
\includegraphics[width=3.0in]{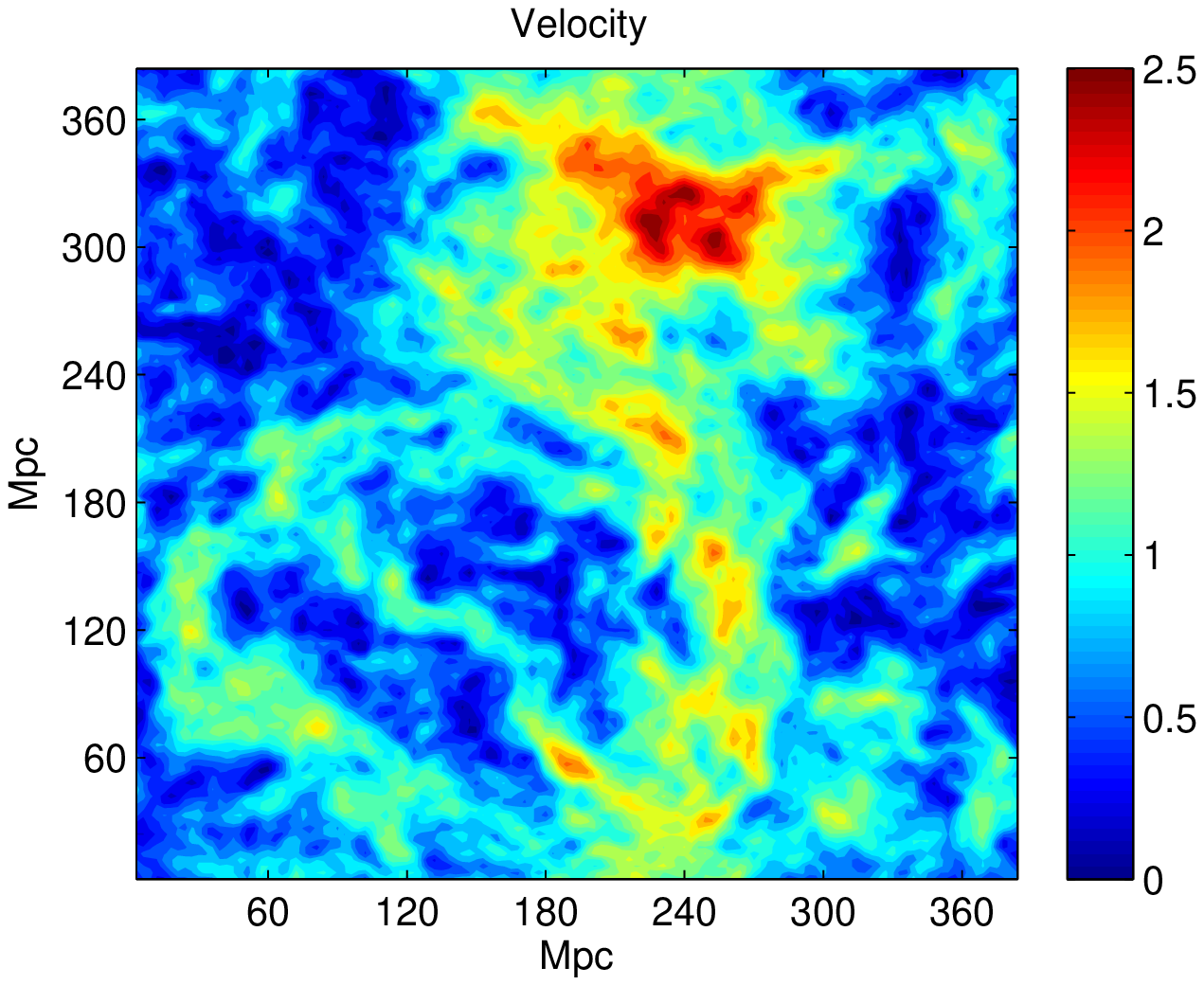}
\includegraphics[width=3.0in]{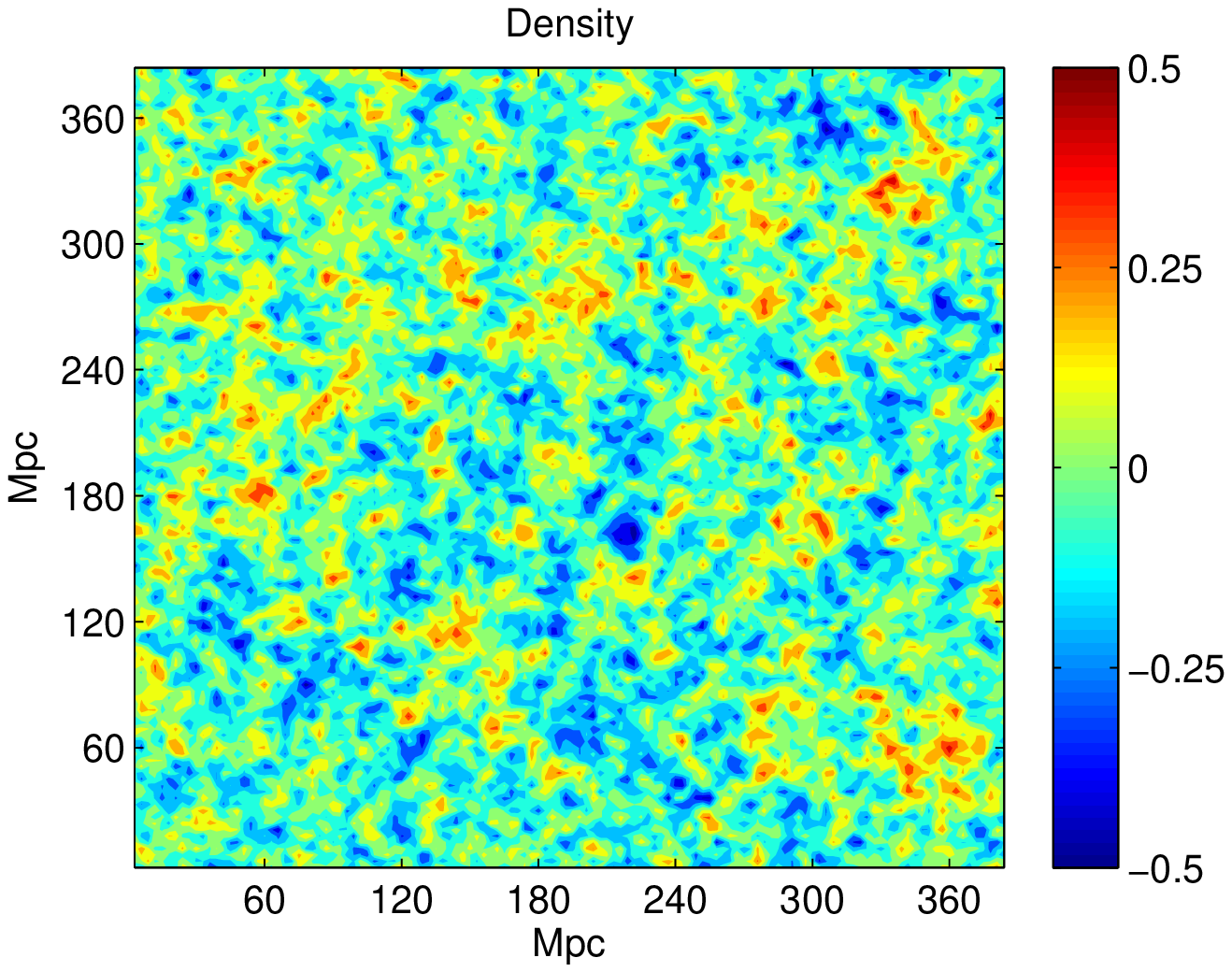}
\includegraphics[width=7.2in]{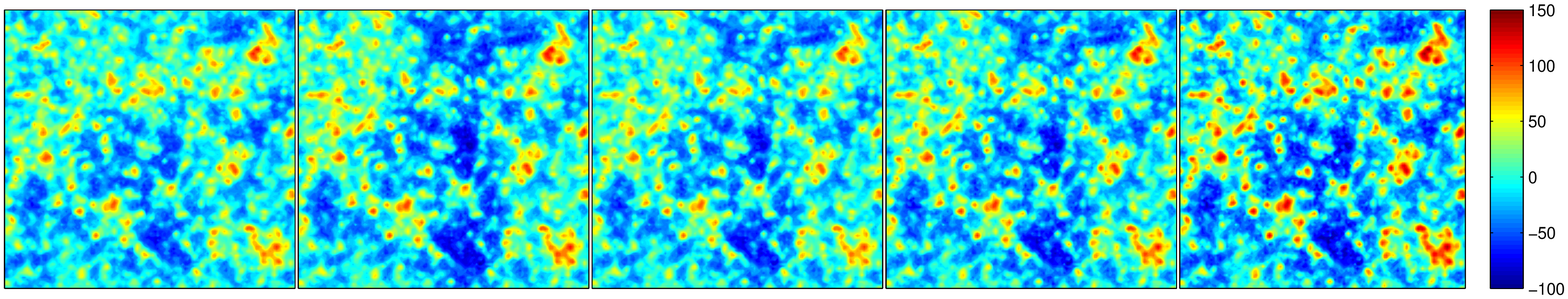}
\caption{A two-dimensional slice of our simulated volume. All panels
show the same slice, i.e., with the same initial conditions. {\bf
Top:} Left: The magnitude of the relative velocity between baryons and
dark matter shown in units of the velocity's root-mean-square
value. Right: The relative fluctuation in density at redshift 20. {\bf
Bottom:} The 21-cm brightness temperature $T_{\rm b}$ (relative to the
cosmic mean in each case) in mK, shown at $z_0+3$. The cases shown
(from left to right) are: no feedback, no $\vbc$; no feedback; weak
feedback; strong feedback; and saturated feedback (where the last four
include the streaming velocities). The cosmic mean values (which have
been subtracted from the maps) are: $\langle T_{\rm b} \rangle = -81$
mK, -89 mK, -73 mK, -88 mK and $-108$ mK, respectively.  }
\label{fig:maps}
\end{figure*}

The most interesting 21-cm signature of the first stars is the
enhanced large-scale fluctuation level due to the supersonic streaming
velocity. A typical two-dimensional slice of our simulated volume is
shown in Figure~\ref{fig:maps} together with the spatial distribution
of the fluctuations in the density and in the relative velocity. A
snapshot of the Universe at a fixed redshift would look very
distinctive in the various feedback cases mainly due to the overall
delay in the heating due to the change in the mean heating rates. It
is more instructive to compensate for this shift and instead compare
each case to the others at the same time relative to the individual
heating redshift (we use $z_0$).

As seen in Figure~\ref{fig:maps}, the feedback weakens the effect of
the relative motion, since it boosts the minimal cooling mass so that
the stars form in heavier halos which are less sensitive to the
relative velocities. Thus, in Figure~\ref{fig:maps} the 21-cm signal
with saturated feedback shows the same pattern as in the no $\vbc$
case (with no feedback); namely, both of them follow the density
fluctuations but with a strong enhancement, where the bias is stronger
for the saturated feedback (since more massive halos are more strongly
biased). The no feedback case (with $\vbc$) shows a strong imprint of
the velocity field along with the influence of density fluctuations;
e.g., the enhanced $T_{\rm b}$ at the bottom right is mainly due to a
density enhancement, while the void at the top (just right of center)
is mainly due to a large relative velocity (but note that the 21-cm
maps are the result of a three dimensional calculation of the
radiation fields, so they cannot be precisely matched with
two-dimensional slices of the density and velocity fields).

The two realistic feedback cases are intermediate, showing a clear
velocity effect, though not as strong as in the no-feedback case. To
understand the comparison between the weak and strong feedback, we
note that the velocities cause a very strong suppression of star
formation up to a halo mass $M\sim 10^6 M_{\odot}$, but above this
critical mass the suppression and its $M$-dependence weaken
considerably (Figure~\ref{fig:dfdm}, right panel). Thus, once the LW
feedback passes through its central redshift, the remaining $\vbc$
effect changes only slowly with $M$, so that around the time of the
heating transition, the weak and strong feedback cases show a similar
fluctuation pattern. However, the strong dependence of bias on $M$
remains, so that the strong feedback case leads to larger fluctuations
on all scales.

\begin{figure*}
\includegraphics[width=3.4in]{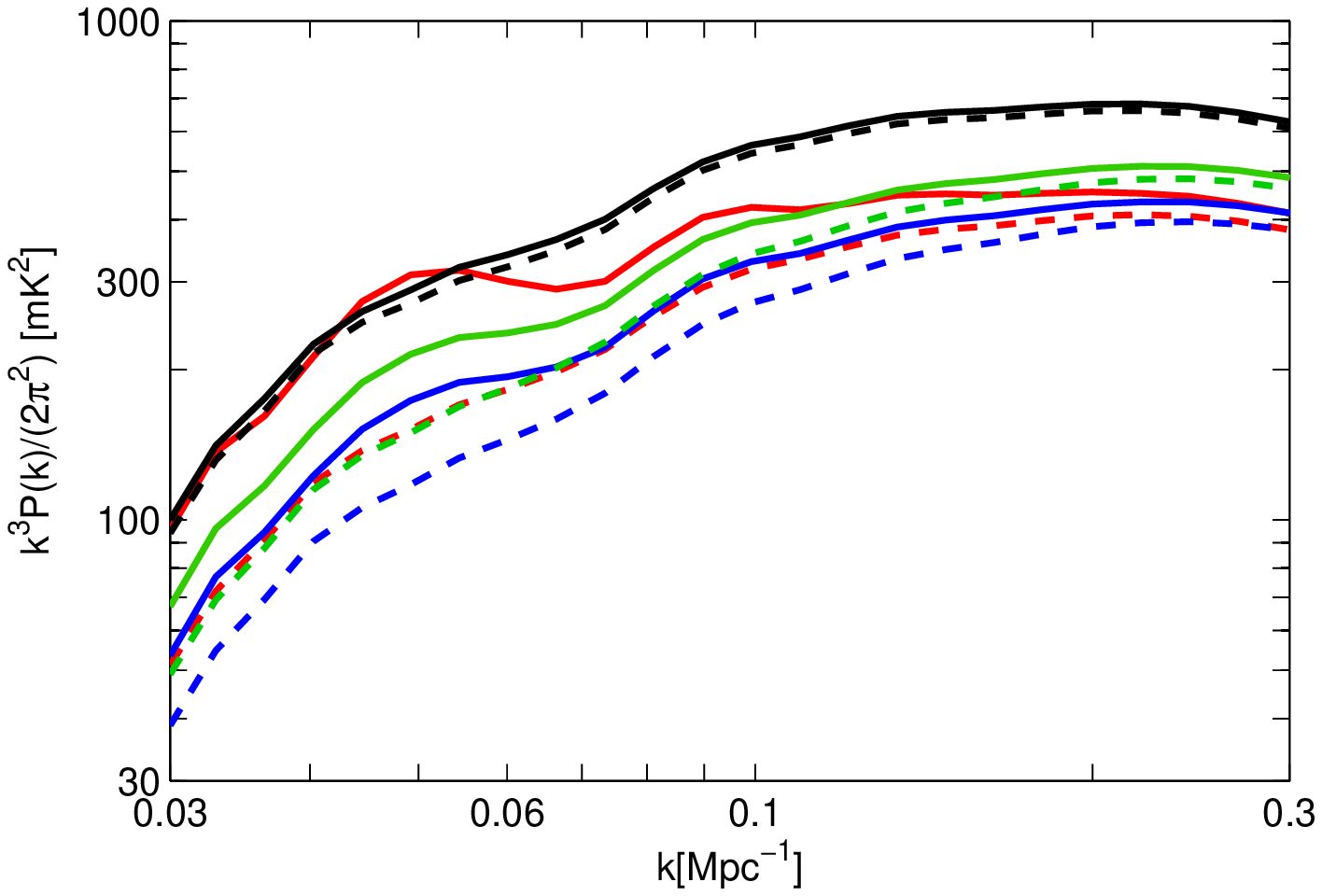}
\includegraphics[width=3.4in]{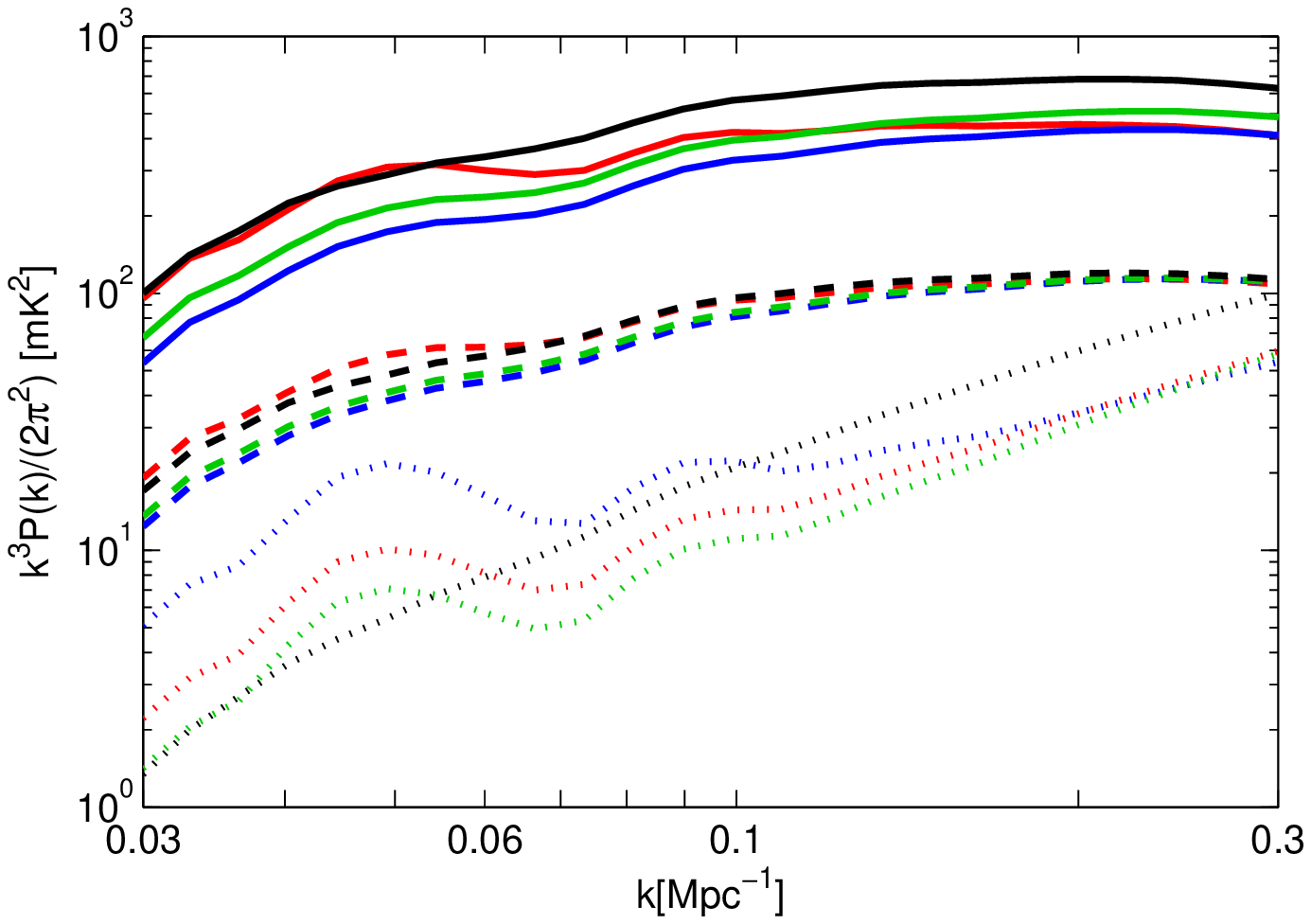}
\caption{Power spectra of the 21-cm brightness temperature for
no feedback (red), weak feedback (blue), strong feedback (green) and
saturated feedback (black). {\bf Left}: $z = z_0 + 3$ with (solid) or
without (dashed) the relative velocity. {\bf Right}: Including $\vbc$,
at redshifts $z = z_0$ (dashed), $z = z_0 + 3$ (solid) and $z = z_0 +
9$ (dotted).}
\label{fig:ps}
\end{figure*}

These and other features can be seen more clearly and quantitatively
in the power spectra (Figure~\ref{fig:ps}). The 21-cm fluctuations
initially rise with time as the heating becomes significant (first in
the regions with a high stellar density). Eventually, as the heating
spreads, the 21-cm fluctuations decline, since the 21-cm intensity
becomes independent of the gas temperature once the gas is much hotter
than the CMB (Equation~\ref{eq:Tb}). Thus, the power spectrum reaches
its maximum height somewhat earlier than $z_0$. The comparison among
the various feedback cases is complex, since the negative LW feedback
has several different effects: 1) The lowest-mass halos are cut out,
reducing the effect of the streaming velocity; 2) The higher-mass
halos that remain are more highly biased; 3) The overall suppression
of star formation delays the heating and LW transitions to lower
redshifts; and 4) Since the higher-mass halos that remain correspond
to rarer fluctuations in the Gaussian tail, their abundance changes
more rapidly with redshift, making the heating transition more rapid
(i.e., focused within a narrower redshift interval). Thus, at $z_0+3$
the large-scale ($k=0.05$ Mpc$^{-1}$) peak is lower for the realistic
feedback cases than it would be with no feedback (due to effect \#1),
and higher for strong feedback than for the weak case (due to effect
\#2). Further back in time ($z_0+9$), weak feedback gives a higher
large-scale peak than both strong feedback (due to effect \#4) and no
feedback (due to effect \#2); at that redshift, saturated feedback
shows no velocity effect (due to effect \#4).

Lower redshifts offer improved observational prospects, due to the
lower foreground noise, which makes negative feedback advantageous due
to effect \#3, above. We find that the most promising redshift is
$z\sim z_0+3$ (Table~1). Assuming a first-generation radio telescope
array with a noise power spectrum that scales as $(1+z)^{5.2}$
\citep{McQuinn:2006, Visbal:2012}, the maximal S/N of the large-scale
($k=0.05$ Mpc$^{-1}$) peak is 3.24 for weak feedback (at $z=18.3$) and
3.91 for strong feedback (at $z=17.7$). For comparison, the
no-feedback case considered in \citet{Visbal:2012} gave (at
$z=z_0=20$) a S/N of only 2.0~. It would be particularly exciting to
detect the evolution of the 21-cm power spectrum throughout the
heating transition, as we suggested in \citet{Visbal:2012}. The S/N
remains above unity at all $z < z_0+7.9 = 23.1$ in the case of the
weak feedback and $z < z_0+7.2 = 21.9$ for the strong feedback (down
to $z=10$ where our simulations end). The streaming velocity clearly
plays a key role in creating this extended observable redshift range,
by boosting the large-scale power (Figure~\ref{fig:ps}).

\begin{table*}
\label{Tab:SN}
\begin{center}
\begin{tabular}{ |l ||c |  c | c |c | c || c | c| c | c |r | }
\hline
\multirow{2}{*}{$z-z_0$}&\multicolumn{5}{|c||}{$\delta T_{\rm b}
(k=0.05~{\rm Mpc}^{-1})$, S/N} &\multicolumn{5}{|c|}{BAO, S/N} \\
 & no $\vbc$ & no fbk & weak & strong & sat & no $\vbc$ & no fbk &
 weak & strong & sat \\
\hline
$-3$& 1.07 & 1.24  & 1.60 & 1.73 & 1.84 & 0.45 & 0.58 &  0.72 & 0.76 & 0.79\\
$0$& 1.68 & 2.33  & 2.35 & 2.69 & 3.09& 0.70& 1.26 &  1.16 &1.30 & 1.31\\
$3$ & 2.26 & 3.59  & 3.24 & 3.91 & 4.74 &0.91 & 2.18&  1.79 & 2.14 &2.00 \\
$6$ & 1.02 & 1.75 & 2.08 & 1.89 & 1.34 & 0.37& 1.17& 1.30&  1.18 & 0.54 \\
$9$ & 0.086 & 0.33  & 0.56 & 0.34& 0.31 & 0.051&  0.23  & 0.41 &0.25 & 0.14 \\
$12$ & 0.18& 0.23  & 0.24 & 0.27 &0.34&  0.083 & 0.099 & 0.11 &0.12& 0.15 \\
\hline
\end{tabular}
\caption{The signal to noise ratio S/N (i.e., the square root of the
ratio between the power spectra of the signal and noise), for a
projected first-generation radio array. We show the S/N of the
large-scale peak at the wavenumber $k=0.05$ Mpc$^{-1}$ (Left), and of
the BAO component (Right), at various redshifts, for five cases: no
feedback no $\vbc$. no feedback (with $\vbc$), weak, strong and
saturated feedback. The BAO S/N is defined as the square root of the
difference between the peak at $k=0.05$ Mpc$^{-1}$ and the trough at
$k=0.07$ Mpc$^{-1}$, each measured with respect to the non-BAO power
spectrum (i.e., the power spectrum smoothed out using a quartic fit),
and each normalized by the noise power spectrum at the same $k$ at the
corresponding redshift.}
\end{center}
\end{table*}

We suggested in \citet{Visbal:2012} that beyond just detecting the
power spectrum, it would be particularly remarkable to detect the
strong BAO signature, since this would confirm the major influence of
the relative velocity and the existence of small ($10^6 M_\odot$)
halos. We find that the S/N for the large-scale BAO feature of the
power spectrum is typically $\sim 0.5 - 0.7$ times that of the
large-scale peak itself (Table~1). In particular, the BAO S/N also
peaks at $z_0+3$, exceeds unity at $z_0-0.7 <z < z_0+ 6.9$ (weak) and
$z_0-1.1 < z < z_0+ 6.4 $ (strong) and reaches a maximum value of 1.79
(weak) or 2.14 (strong feedback).

We have assumed here the projected sensitivity of a thousand-hour
integration time with an instrument like the Murchison Wide-field
Array \citep{MWAref} but designed to operate in the range of 50--100
MHz. An instrument similarly based on the Low Frequency Array
\citep{LOFARref} should improve the S/N by a factor of $\sim 1.5$,
while a second-generation instrument like the SKA or a 5000-antenna
MWA should improve it by at least a factor of 3 or 4
\citep{McQuinn:2006, Visbal:2012}. Thus, future instruments may be
able to probe even earlier times, including the central stages of the
LW feedback.

\section{Conclusions}

We have presented new predictions for the signature of the first stars
in the heating fluctuations of the 21-cm brightness temperature. We
ran hybrid simulations that allow us to predict the large-scale
observable 21-cm signature while accounting on small-scales for
various effects on star-formation investigated by previous analytical
models and numerical simulations. In particular, we incorporated for
the first time the Lyman-Werner feedback on star formation, calculated
self-consistently including the effect of the supersonic streaming
velocity. A three-dimensional calculation of the Lyman-Werner and
X-ray backgrounds allowed us to calculate the heating history of the
gas and the resulting 21-cm intensity maps.

We have focused on the negative LW feedback, which begins at $z \sim
30-40$ but strengthens very gradually, passing its central point at
$z_\LW \sim 19-24$ and saturating only at $z \sim 10$. The heating
transition is centered at $z_0 \sim 15$ (including a delay of $\Delta
z \sim 1.5-2$ due to the feedback), when the LW transition is well
advanced but still far from saturated. The large-scale 21-cm power
spectrum is potentially observable over a broad redshift range of $z
\sim 10-22$ or 23. The best prospects are at $z \sim 18$, when the
large-scale peak reaches a signal-to-noise ratio (for a projected
first-generation radio telescope array) of 3.2 (for our weak feedback
case) or 3.9 (for strong feedback). At this redshift, the BAO
signature (which marks the velocity effect and the presence of $10^6
M_\odot$ halos) should also be observable with a S/N$\sim 2$. The BAOs
should be observable over a broad redshift range of $\Delta z \sim
7.5$~.

These numbers are obtained with our standard set of expected
astrophysical parameters, but they may shift around a bit depending on
the precise properties of the early stars and their remnants. We hope
these findings will stimulate additional numerical simulations of the
complex radiative feedback at $z \sim 10-30$, as well as future
observational efforts in 21-cm cosmology directed at the epoch prior
to reionization.

\section{Acknowledgments}
We are most grateful to Zoltan Haiman for supplying us with the atomic
data on the Lyman-Werner absorption lines. This work was supported by
Israel Science Foundation grant 823/09. A.F.\ was also supported by
European Research Council grant 203247.


\label{lastpage}

\end{document}